\documentclass[12pt,preprint]{aastex}

\shorttitle{Panoramic Opacity Mapping}
\shortauthors{Shamir \& Nemiroff}

\begin{document}

\title{All-sky Relative Opacity Mapping Using Night Time Panoramic Images}

\author{Lior Shamir}
\affil{Department of Physics, Michigan Technological University, Houghton, MI 49931}
\email{lshamir@mtu.edu}

\and

\author{Robert J. Nemiroff}
\affil{Department of Physics, Michigan Technological University, Houghton, MI 49931}

\begin{abstract}
An all-sky cloud monitoring system that generates relative opacity maps over many of the world's premier astronomical observatories is described.  Photometric measurements of numerous background stars are combined with simultaneous sky brightness measurements to differentiate thin clouds from sky glow sources such as air glow and zodiacal light. The system takes a continuous pipeline of all-sky images, and compares them to canonical images taken on other nights at the same sidereal time.  Data interpolation then yields transmission maps covering almost the entire sky. An implementation of this system is currently operating through the Night Sky Live network of CONCAM3s located at Cerro Pachon (Chile), Mauna Kea (Hawaii), Haleakala (Hawaii), SALT (South Africa) and the Canary Islands (Northwestern  Africa). 
\end{abstract}

\keywords{atmospheric effects -- methods: data analysis}

\section{Introduction}

The effectiveness of ground-based telescopes is modulated by atmospheric opacity.  Previously implemented systems created to monitor clouds and sky opacity have been limited to monitoring a single point, usually the zenith \citep{Ana04,Cha03}. Single-point opacity measurements, however, inherently ignore even dramatic non-uniform opacity across the night sky. Information regarding the opacity of the entire sky can be used in a variety of situations. For example, a robotic telescope surveying the entire sky would be used inefficiently if it observed a cloudy part of the sky, while other parts of the sky remained relatively clear. Human observers also benefit from using these data by manually selecting from equally desirable targets based on local sky transparency (SpaceWatch, private communication).

A cloud monitoring system utilizing the thermal infrared and covering 135 degrees$^2$ operates at Apache Point \citep{Hog01,Hul94}.  The Apache Point system gives a simple boolean indication of the existence of a cloud at a certain point in the sky.  The system is used to accept or reject collected data.

Providing a real-time fisheye view of the optical night sky, the {\it Night Sky Live} (NSL) network of CONtinuous CAMeras (CONCAMs) automatically creates a pipeline of all-sky images that is routinely used by astronomers as a contemporary cloud monitor \citep{Rut03,Per00,Per05}.  NSL CONCAM3 images are slightly more sensitive than the human eye, particularly in the CCD-sensitive red part of the spectrum.  This allows sub-visual cirrus clouds, airglow, and other sky glows such as zodiacal light to be easily discernable on the images, along with over a thousand stars.  This NSL network and its opacity mapping software is only one realization of a more general visual night sky monitoring system that can generate opacity maps.  In this paper, we therefore first delineate the workings of a general night sky opacity mapping system, and later describe our specific NSL CONCAM implementation. In Section~\ref{canonical_frames} we describe the construction of the database of canonical images, in Section~\ref{opacity_maps} we describe the process in which opacity maps are generated, in Section~\ref{accuracy} we discuss the accuracy of the system and in Section~\ref{nsl} we discuss our implementation of the sky opacity monitoring system using the Night Sky Live network.

\section{Canonical Image Database}
\label{canonical_frames}

A first step to building an opacity map is to create a comparison image from which any image can be compared.  This image can be composed of images taken at the same sidereal time as the given frame, and will be called the ``canonical image''.  One can think of the ``canonical image'' as a hypothetical image taken when the sky is completely clear.  In reality, the canonical image is better created from several relatively clear images taken at the same sidereal time.
To proceed, the system first needs to acquire a sufficient database of relatively clear images.

One approach of obtaining this database is to manually select images taken at clear nights. However, observing images by eye is a time-consuming task, and the required human resources are not always available. Moreover, since thin cirrus clouds are hard to detect, a human observer might consider some images as {\it clear} while in fact parts of the sky are covered with light clouds.

Another approach, which we have found more appropriate, is to use an algorithm that automatically searches for images taken at clear nights. The algorithm is based on applying a star recognition algorithm that associates the point spread functions in the image with star catalog entries \citep{Sha05}. After the point spread functions are associated with known stars, the algorithm checks if the stars that are brighter than a pre-set visual magnitude are detected at a pre-set statistical confidence level.  If the PSFs of a sufficient percentage of expected stars are found in the image, the image is classified as {\it clear}, and added to the database.  

The percentage of stars visible at a certain frame is dependent not only on the sky clarity, but also on the darkness of the sky. For instance, faint stars near the galactic plane will not be detected as easily as stars far from the Milky Way. Therefore, the pre-set threshold of visual magnitude should be chosen such that stars near the center of the galactic plane brighter than the threshold can be detected.  Since this problem becomes more substantial when the moon is up, canonical images are added to the database only when the moon is down.

When an image is added to the database, it is averaged with all other images taken at the same sidereal time that are already present. Averaging the images gives a better signal to noise ratio than using a single exposure. The large pixel size and the absence of moving parts in the all-sky monitoring system simplify the process of image co-adding.

The algorithm described above is used in order to build the database while the system is operating by adding all images that are classified by the algorithm as {\it clear}. This policy allows the system to provide useful data even in the first days of its operation, while autonomously improving the quality of the data with time.

An alternative approach would be to iterate toward a standard canonical image that does not change after a given number of images are averaged.  Once the standard canonical image is established, it is understood that all opacity measurements are given relative to this standard.

\section{All-Sky Relative Opacity Maps}
\label{opacity_maps}

The building of relative opacity maps requires several logical steps.  The first is the building of a canonical image taken from previous images taken at the same sidereal time, as discussed last section.  Another step is the rejection of pixels dominated by cosmic-ray generated counts \citep{Sha05b}.  Such rejections might make use of the non-point source nature of cosmic ray splashes, or the fact that the apparent cosmic ray ``source" cannot be found on the canonical frame.  Next, bright planets and variable stars with relatively large amplitude such as Algol \citep{Muz05} are also rejected using a star recognition algorithm such as \citep{Sha05}.  Any PSF that is a given $\sigma$ brighter than its background is assigned with two values: a stellar intensity, and an estimated intensity of the background. The values assigned to the PSFs are then compared with the values of the PSFs of a canonical image. The comparison is based on Equations~\ref{star_glow} and~\ref{sky_glow}.

\begin{equation}
\label{star_glow}
M_{\mathrm star} = T \cdot I_{\mathrm star} + T \cdot I_{\mathrm space} + I_{\mathrm cloud}
\end{equation}

\begin{equation}
\label{sky_glow}
M_{\mathrm background}= T \cdot I_{\mathrm space} + I_{\mathrm cloud}
\end{equation}

Where $M_{\mathrm star}$ is the intensity of the light coming from pixels at the location of a star in the given image, $M_{\mathrm background}$ is the intensity of the light coming from pixels {\it not} at the location of a star, $I_{\mathrm star}$ is the intensity of the star, $I_{\mathrm space}$ is the intensity of light from background space, $I_{cloud}$ is the intensity of the cloud covering the star and $T$ is the transmission around the star. Using Equations~\ref{star_glow} and~\ref{sky_glow}, the transmission is determined by Equation~\ref{sky_opacity}.

\begin{equation}
\label{sky_opacity}
T=\frac{(M_{\mathrm star} - M_{\mathrm background})}{I_{\mathrm star}} 
\end{equation}

Where $I_{\mathrm star}$ is based on measurements taken from the canonical image such that $I_{\mathrm star}=Mo_{\mathrm star}-Mo_{\mathrm background}$, where $Mo_{star}$ and $Mo_{background}$ are the measured intensity of the star and the background in the canonical image. This calculation of $I_{\mathrm star}$ is based on the assumption that no light is lost as starlight travels through air on a clear night. This assumption, however, is not true, and provides a simplification of the problem that allows estimating the normalized transmission comparing to a clear night. This simplification gives Equation~\ref{sky_opacity1}.

\begin{equation}
\label{sky_opacity1}
T=\frac{(M_{\mathrm star} - M_{\mathrm background})}{(Mo_{\mathrm star} - Mo_{\mathrm background})} 
\end{equation}

Since the desired product is a broad relative transmission map, the transmission of each pixel is calculated by interpolating the transmission measured directly along the lines to the stars in the image.  This is performed by choosing, for example, the four nearest stars $S_{left}, S_{top}, S_{right}, S_{bottom}$ such that $|Y_{Sl}-Y_0|<X_0-X_{Sl}$, $|Y_{Sr}-Y_0|<X_{Sr}-X_0$, $|X_{St}-X_0|<Y_0-Y_{St}$, $|X_{Sb}-X_0|<Y_{Sb}-Y_0$, where $X_{Sn}$ is the X image coordinate of the star $S_n$, $Y_{Sn}$ is the Y image coordinate of the star $S_n$, and $(X_0,Y_0)$ are the image coordinates of the given pixel that its sky transmission is being estimated.  After finding the four nearest stars, a two-dimensional linear interpolation of the transmission measured by the four stars is performed, and the calculated value is determined as the relative transmission of the pixel. The computed relative transmission values of all pixels in the frame are then used for generating the all-sky opacity maps such that high relative transmission is an indication of high relative sky opacity, while low relative transmission indicates low relative opacity.

The computed normalized transmission $T$ provided by this method is not intended to be used for scientific purposes such as photometric reduction. However, providing a numeric indication of the estimated atmospheric transmission is believed to have an advantage over a simple boolean clear/cloudy bit of information. The numeric indication can be used by robotic telescopes for the purpose of decision making (for instance, weighing the importance of the observation against the atmospheric transmission) and can also provide human observers with more informative indications regarding the visibility conditions of the area of the sky they observe.

\section{System Accuracy}
\label{accuracy}

The maps generated by the presented system provide an approximation of the relative transmission of the sky. The accuracy of the data is dependent on the noise of the CCD measurements. When the gain is 1, the estimated error of the measured intensity of the star is defined by Equations~\ref{c_error} and~\ref{c0_error}.

\begin{equation}
\label{c_error}
\sigma_{\mathrm star}=\sqrt{M_{\mathrm star}}
\end{equation}

\begin{equation}
\label{c0_error}
\sigma_{\mathrm star0}=\sqrt{Mo_{\mathrm star}}
\end{equation}

Assuming a perfect detector, when the intensity of the background $M_{\mathrm background}$ is calculated by averaging the 1600 pixels around the peak of the PSF, the estimated error of the background is defined by Equations~\ref{b_error} and~\ref{b0_error}

\begin{equation}
\label{b_error}
\sigma_{\mathrm background}=\frac{M_{\mathrm background}}{\sqrt{1600}}
\end{equation}

\begin{equation}
\label{b0_error}
\sigma_{\mathrm background0}=\frac{Mo_{\mathrm background}}{\sqrt{1600}}
\end{equation}

Let $\sigma_{\Delta}$ and $\sigma_{\Delta0}$ be the estimated error of $(M_{\mathrm star}-M_{\mathrm background})$ and $(Mo_{\mathrm star} - Mo_{\mathrm background})$ respectively. $\sigma_{\Delta}$ and $\sigma_{\Delta0}$ are defined by Equations~\ref{error_delta} and~\ref{error_delta0}.

\begin{equation}
\label{error_delta}
\sigma_{\Delta}=\sqrt{\sigma_{\mathrm star}^2+\sigma_{\mathrm background}^2}
\end{equation}

\begin{equation}
\label{error_delta0}
\sigma_{\Delta0}=\sqrt{\sigma_{\mathrm star0}^2+\sigma_{\mathrm background0}^2}
\end{equation}

The fractional standard deviation of the transmission $T$ (calculated by Equation~\ref{sky_opacity1}) is defined by Equation~\ref{tau_error}.

\begin{equation}
\label{tau_error}
\sigma_{T} = \sqrt{
({\sigma_{\Delta} \over {M_{\mathrm star} - M_{\mathrm background}}})^2+
({\sigma_{\Delta0} \over {Mo_{\mathrm star} - Mo_{\mathrm background}}})^2
}
\end{equation}

For example, suppose that $M_{\mathrm star}=6000$, $Mo_{\mathrm star}=7000$ and $M_{\mathrm background}=Mo_{\mathrm background}=1000$. Using Equation~\ref{sky_opacity1}, $T$ is $\sim$0.83, giving $\Delta M=-2.5\log0.83\simeq0.2$. Using Equations~\ref{c_error} -~\ref{b0_error}, $\sigma_{\mathrm star}\simeq77.46$, $\sigma_{\mathrm star0}\simeq83.67$ and $\sigma_{\mathrm background}=\sigma_{\mathrm background0}=2.5$. Equations~\ref{error_delta} and~\ref{error_delta0} give $\sigma_{\Delta}\simeq77.5$ and $\sigma_{\Delta0}\simeq83.71$, and Equation~\ref{tau_error} gives  $\sigma_{T}\simeq0.021$. Therefore, $\sigma_{\Delta M}=-2.5\log({{0.83-0.021\cdot0.83} \over 0.83})\simeq0.023$.

\begin{figure}
\includegraphics{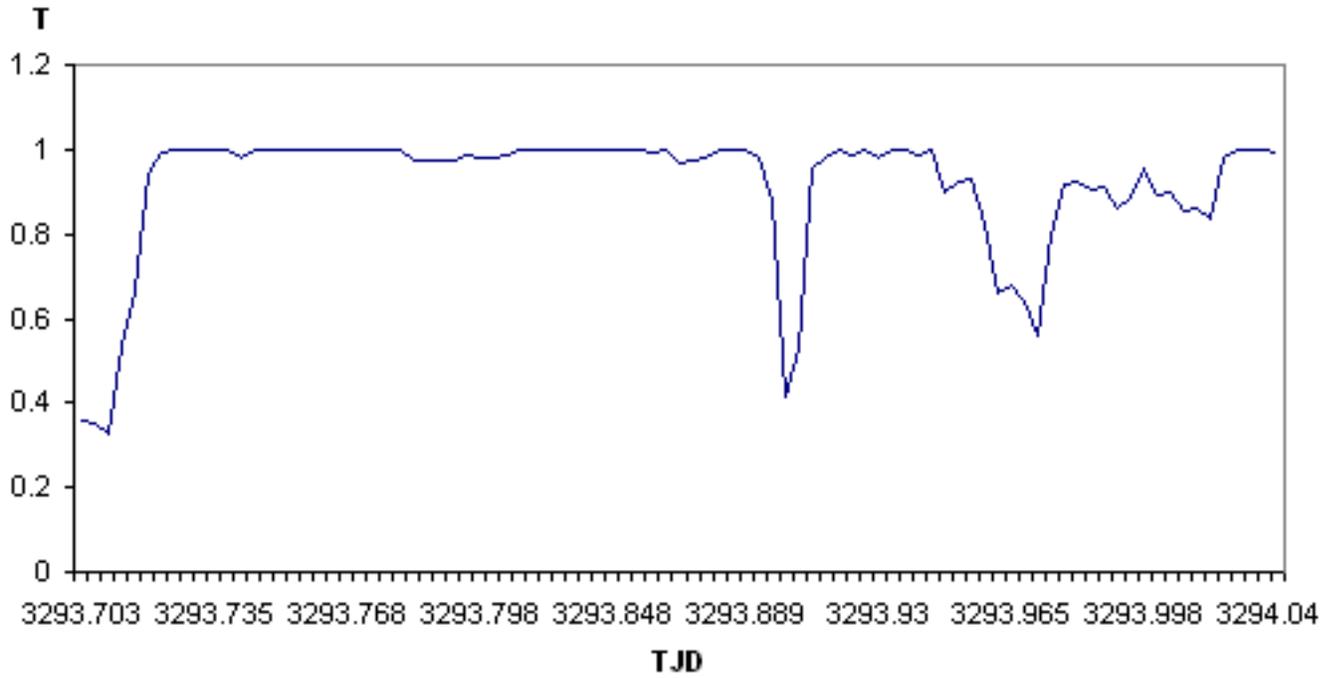}
\caption{Transmission measured at RA=00h 40m, DEC=56.5$^{o}$ recorded at Mauna Kea on 10/15/2004}
\label{night_measure}
\end{figure}

\begin{figure}
\includegraphics{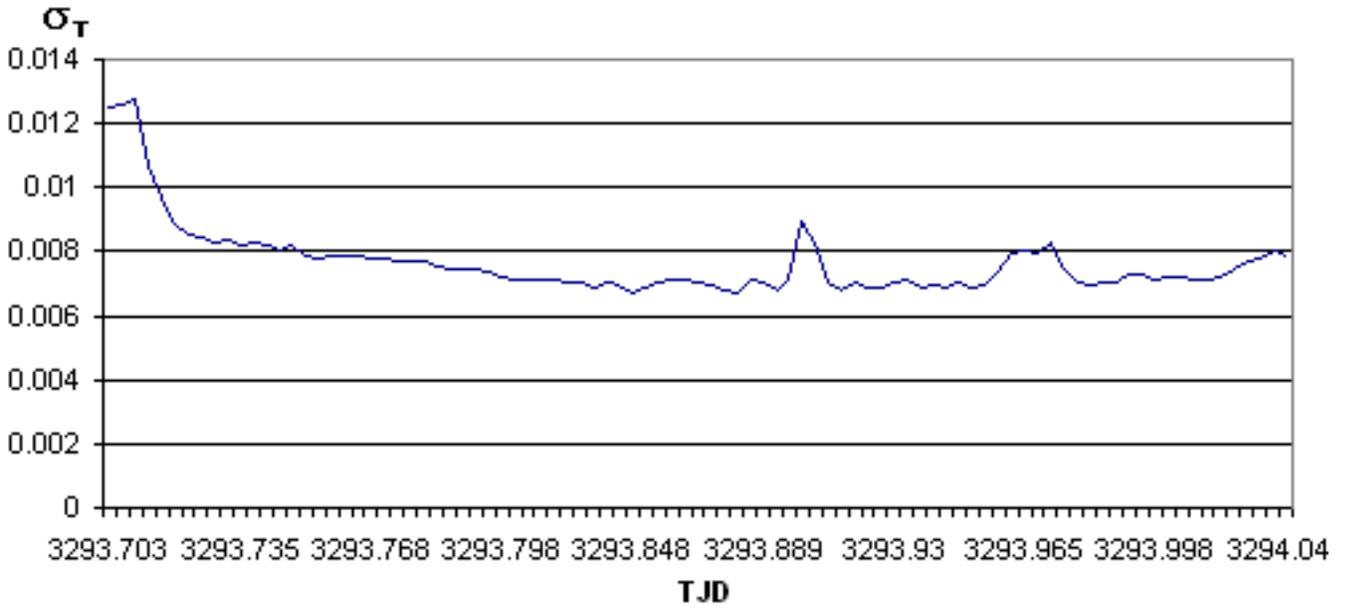}
\caption{$\sigma_{T}$ of RA=00h 40m, DEC=56.5$^{o}$ recorded at Mauna Kea on 10/15/2004}
\label{night_measure_error}
\end{figure}

Figure~\ref{night_measure} shows transmission measurements at RA=00h 40m, DEC=56.5$^{o}$ taken using the {\it Night Sky Live} implementation of the proposed method described in Section~\ref{nsl}. The data were recorded at Mauna Kea throughout the night of 10/15/2004. Figure~\ref{night_measure_error} shows the calculated values of $\sigma_{T}$.

\begin{figure}
\includegraphics{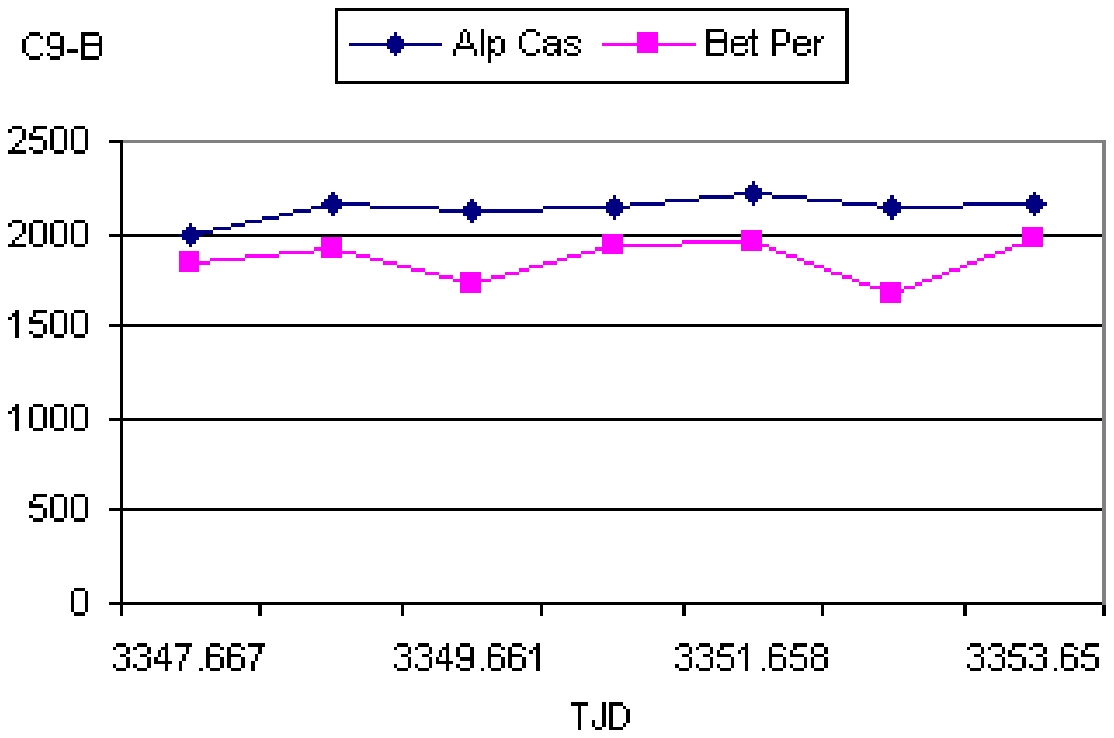}
\caption{The average of the 9 brightest pixels of Alp Cas and Bet Per subtracted by the background, as measured by {\it CONCAM} at KPNO between 12/8/2004 and 12/15/2004}
\label{algol_alpcas}
\end{figure}

One problem that may affect the reliability of the system is stellar variability.  All-sky monitoring devices such as the {\it CONCAM}s \citep{Nem99} are sensitive to variable stars \citep{Muz05}, and in some cases can sense the variability of low-amplitude variable stars such as Polaris \citep{Nem05}. An eclipsed star, for example, might create a false region of high opacity.  Figure~\ref{algol_alpcas} shows the measurements of a variable star (Algol) and a non-variable star (Alpha Cassiopeia) over the period between 12/8/2004 and 12/15/2004. The data were recorded at KPNO using an implementation of the proposed system described in Section~\ref{nsl} by taking measurements on different days at the same sidereal time. The graph shows the average of the 9 brightest pixels of the PSFs of the two stars subtracted by the estimated local background. The decreased luminosity of Algol during the eclipse might mistakenly lead the system to conclude that Algol is covered by a cloud. However, since known variable stars with amplitudes larger than a given amount are ignored, the variability of Algol is not expected to have any effect on the results.  Additionally, each pixel is assigned with a sky transmission value based on more than one star, further minimizing variability error.

Another problem affecting the reliability of the system is transmission local to the optics that has nothing to do with the real transmission of the sky.  The most common example is water droplets on the lens, although even spiders have been noted in this regard.  The result, of course, is an artificially lower $I_{\mathrm star}$, and therefore an artificially low transmission estimate for part of the sky.  Effects like these are sometimes unavoidable, however experience shows that a heated glass dome protecting the lens can be a practical deterrent.

One method of detecting the presence of moisture is by comparing several images. If the measured opacity is always lower at the same region of the sky (in terms of topocentric coordinates), it is likely that the lens is covered with moisture in that region. A simpler approximation of the problem is to count the number of stars brighter than a certain threshold as described in Section~\ref{canonical_frames}. If the percentage of stars detected at a certain region of the sky is less than a pre-set low value while other regions are relatively clear, the region is assumed by the system to be covered with moisture.

\section{Implementation Using the {\it Night Sky Live} Network}
\label{nsl}

The technique described in sections~\ref{canonical_frames} -~\ref{opacity_maps} has been implemented using the infrastructure of the Night Sky Live network \citep{Nem99,Nem06}, which consists of 10 nodes called {\it CONCAM} located at some of the world's premier observatories.  Currently CONCAMs operate at the following observatories: Haleakala (Hawaii), Mauna Kea (Hawaii), Mt. Wilson (California), Kitt Peak (Arizona), Rosemary Hill (Florida), Cerro Pachon (Chile), Canary Islands (Spain), SALT (South Africa), Wise (Israel), and Siding Spring (Australia).  Each node incorporates an SBIG ST-8 (CONCAM2s) or ST-1001E (CONCAM3s) CCD camera, a Nikon FC-E8 or SIGMA F4-EX 8mm fish-eye lens and an industrial PC. Each {\it CONCAM} takes one 1024$\times$1024 180-second exposure all-sky FITS \citep{Wel81} image every 236 seconds. The FITS files are then transmitted to the main server where they are copied to the public domain and can be accessed at {\url http://nightskylive.net}. FITS frames are stored in the main server for two months, after which they are archived on DVDs and removed from the server, but are still available upon specific request. Each {\it CONCAM} node of the {\it Night Sky Live} network provides one all-sky image every 3 minutes and 56 seconds.

\begin{figure}
\includegraphics{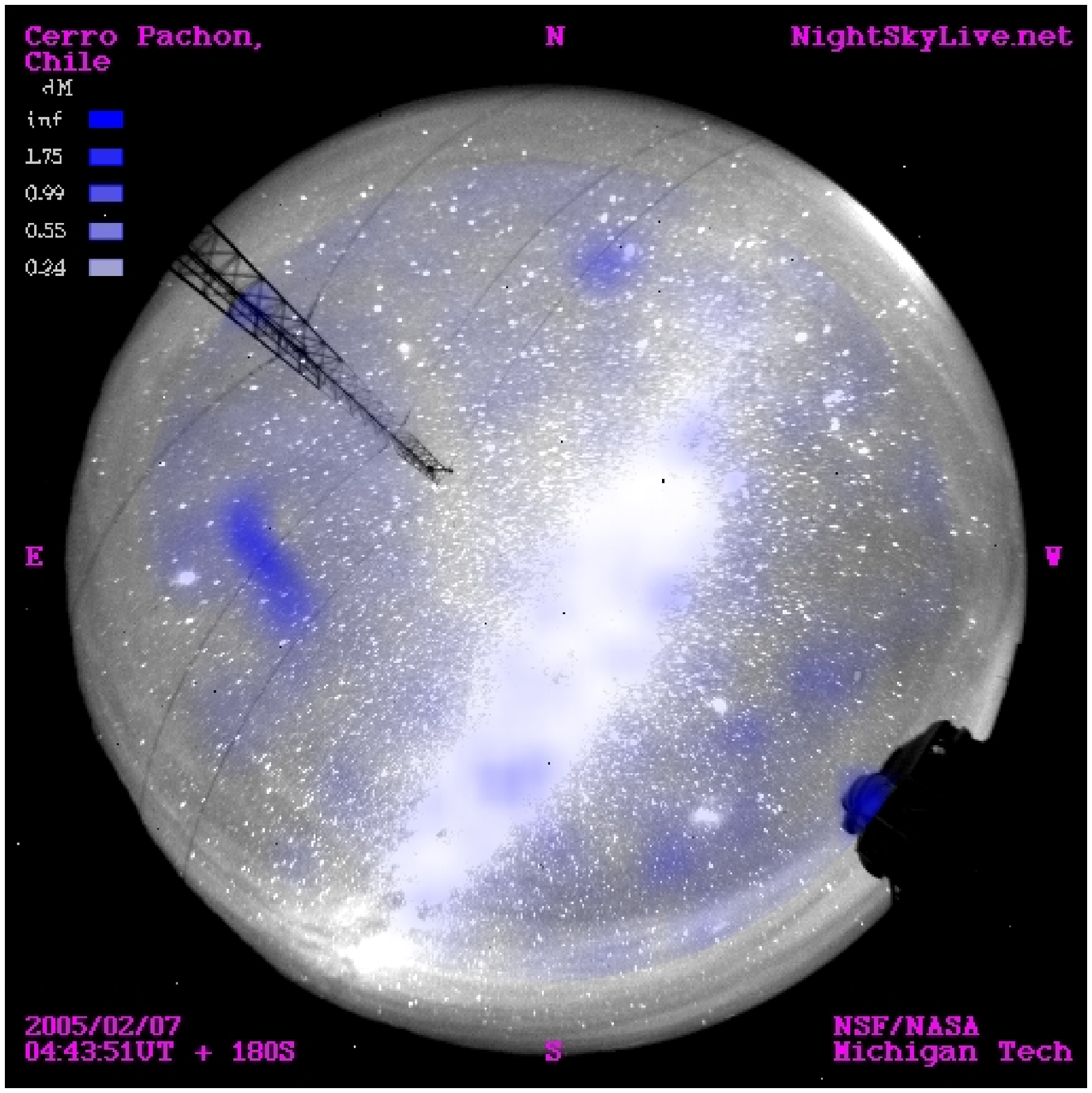}
\caption{All-sky opacity map generated in Cerro Pachon, Chile}
\label{opacity_map}
\end{figure}

The pipeline of all-sky images provides the data required to generate all-sky relative opacity maps using the technique described above.  To date, however, relative opacity maps are only generated for the CONCAM3 stations.  Figure~\ref{opacity_map} is an example of an all-sky opacity map generated by the Cerro Pachon CONCAM3.

In an all-sky image taken in a clear night, CONCAM3s record approximately 1200 stars 20$\sigma$ brighter than their background and at altitude greater than 12$^{o}$ above the horizon. The side of each pixel in a {\it CONCAM} all-sky image is approximately 10$^\prime$, giving that each pixel covers $\sim$100 minutes$^{2}$. Sub-pixel opacity estimation is currently not provided by the system.

In order to visualize the sky opacity map, the pixel is colored in blue such that a stronger blue color represents higher opacity. The maps are generated by changing the B component of each RGB triple such that $B=B_{0}T+255(1-T)$, where $B_0$ is the original B value of the pixel and $T$ is the sky transmission calculated using Equation~\ref{sky_opacity}. A scale added to the top left part of the transmission map presents the colors of several levels of opacity such that $\Delta M = -2.5\log_{10}T$.

In order to test the effectiveness of the data provided by the Night Sky Live implementation of the proposed method, we compared them to data provided by the Mercator 1.2M telescope and its P7-2000 photometer \citep{Ras04} installed at the Roque De Los Muchachos observatory on La Palma, where a CONCAM system is also installed.  The Mercator telescope provides between 10 to 30 single-point atmospheric extinction measurements per night, and the data (including archived data from previous nights) are available on-line.

\begin{figure}
\includegraphics{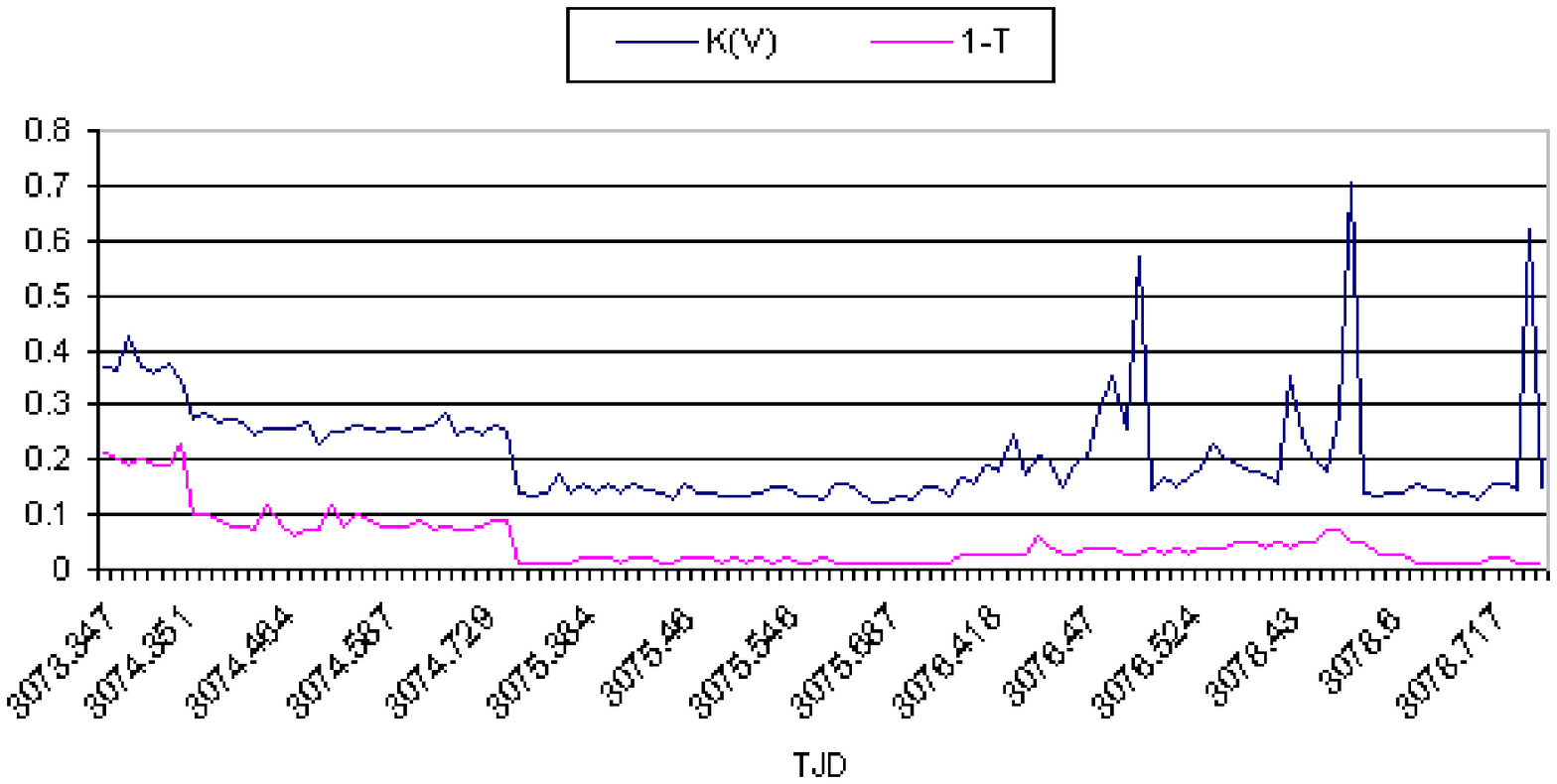}
\caption{K(v) provided by Mercator and 1-T provided by CONCAM measured in La Palma in the period between 3/8/2004 and 3/14/2004}
\label{kv_t}
\end{figure}

Figure~\ref{kv_t} shows a comparison between the K(V) atmospheric extinction measurements provided by Mercator and the transmission computed by the CONCAM implementation of the proposed method in the period between 3/8/2004 and 3/14/2004. Although the data provided by the systems do not provide an exact match, the comparison shows that the two systems are correlated.

\section{Conclusions}

In this paper, a simple cloud monitoring system that provides all-sky relative opacity maps was presented. The maps are generated by comparing all-sky images taken in real-time with canonical images taken on clear nights at the same sidereal time. Although the maps are approximation of the true relative transmission, they have been found useful by observers, and provide data that have never been provided by other available cloud monitoring systems. The system is deployed at five premier observatories, which are Cerro Pachon (Chile), Mauna Kea, Haleakala, SALT (South Africa) and Canary Islands.

\end{document}